# Immersive XR That Moves People:

## How XR Advertising Transforms Comprehension, Empathy, and Behavioural Intention

KOBAYASHI Yuki*,   TOIDA Koichi (MESON, Inc.)

* yuki@meson.tokyo

**Abstract:** Extended Reality (XR) affords an enhanced sense of bodily presence that supports experiential modes of comprehension and affective engagement which exceed the possibilities of conventional information delivery. Nevertheless, the psychological processes engendered by XR, and the manner in which these processes inform subsequent behavioural intentions, remain only partially delineated. The present study addresses this issue within an applied context by comparing non-immersive 2D viewing advertising with immersive XR experiential advertising. We examined whether XR strengthens internal responses to a product, specifically perceived comprehension and empathy, and whether these responses, in turn, influence the behavioural outcome of purchase intention. A repeated-measures two-way ANOVA demonstrated a significant main effect of advertising modality, with XR yielding higher ratings on all evaluative dimensions. Mediation analysis further indicated that the elevation in purchase intention was mediated by empathy, whereas no significant mediating effect was observed for comprehension within the scope of this study. These findings suggest that immersive XR experiences augment empathic engagement with virtual products, and that this enhanced empathy plays a pivotal role in shaping subsequent behavioural intentions.

## 1.  Introduction

Extended Reality (XR) is increasingly characterised not merely as a medium for the presentation or transmission of information, but as a technologically mediated mode of experience that reshapes how individuals perceive and engage with the world. By invoking perceptual processes and spatial self-location, XR emphasises embodied and affective engagement, enabling users to engage with content not passively, as in traditional media, but actively from within an enveloping virtual environment. Such immersive experiences have the potential to recalibrate the psychological stance with which individuals interact with virtual entities. Despite this promise, the psychological transformations brought about by XR experiences, particularly their quantifiable impact in applied contexts, remain only partially understood.

To provide a theoretical lens for interpreting these transformations, the present study draws upon the classical three-component Cognitive-Affective-Behavioural (CAB) model of attitudes (Rosenberg and Hovland, 1960). The CAB model posits that attitudes consist of cognitive elements concerned with understanding and belief formation, affective elements involving empathy and evaluative sentiments, and behavioural elements associated with actions such as purchasing or avoidance. A growing body of empirical work has demonstrated the explanatory reach of the CAB model within XR environments. For example, modelling studies in immersive learning have shown that XR-specific affordances, including presence and agency, influence learning outcomes through both affective determinants, such as interest, motivation, and embodiment, and cognitive determinants, including cognitive load and self-regulation, thereby shaping comprehension (Makransky and Petersen, 2021). Research focusing on the narrative qualities of XR similarly reports that participants who viewed a VR documentary depicting a refugee camp exhibited heightened empathy relative to those who viewed the same material in 2D, with the degree of immersion predicting empathic engagement (Schutte and Stilinović, 2017). In perspective-taking tasks concerning homelessness, VR-based interventions have been shown to increase empathy and prosocial behavioural intentions immediately after exposure, with some effects persisting for up to eight weeks (Herrera *et al.*, 2018).

Applied research in advertising and consumer behaviour, where XR experiences have acquired increasing prominence, likewise draws on the CAB model. Early work on VR advertising indicates that 3D interactive formats improve product knowledge and brand attitudes compared with conventional 2D advertisements, and may in some instances augment purchase intention (Li *et al.*, 2002). Subsequent research has shown that mobile VR brand experiences enhance vividness and telepresence, which strengthen advertising and brand attitudes and, in turn, purchase intention (Van Kerrebroeck *et al.*, 2017). A recent systematic review of consumer research in VR environments notes that evaluative judgements comprise cognitive appraisals such as understanding and perceived realism, as well as affective appraisals including enjoyment, excitement, and emotional engagement, all of which contribute to downstream attitudes and behavioural intentions (Branca *et al.*, 2024). In VR retail contexts, spatial presence and media richness have been found to heighten flow experiences, which subsequently inform cognitive attitudes and empathic judgements, thereby indirectly influencing both online and offline purchasing behaviour (Hsiao *et al.*, 2023). Taken together, these findings suggest that cognitive and affective evaluations, corresponding respectively to comprehension and empathy, may shape behavioural responses such as purchase intention.

Notwithstanding these insights, a considerable proportion of existing studies isolate either comprehension or empathy, or adopt sequential causal accounts in which cognition is presumed to precede affective response and subsequently behaviour. In applied domains, particularly advertising, the psychological mechanisms through which XR influences consumer responses remain insufficiently delineated. Although XR advertising is theoretically expected to deliver a heightened sense of immersion and presence relative to conventional flat-panel formats, empirical investigations remain limited, and the magnitude, contingencies, and mechanisms of such effects are not yet well consolidated (Wedel *et al.*, 2020). In particular, the manner in which XR advertising modulates internal responses such as comprehension and empathy, and the extent to which these responses propagate to behavioural intentions such as purchase intention, requires further empirical scrutiny.

The present study therefore examines XR advertising within an applied experimental paradigm. We compare non-immersive 2D viewing advertising delivered via a flat-panel display with immersive XR experiential advertising delivered through a head-mounted display, and we assess how these modalities influence internal responses, specifically comprehension and empathy, as well as the behavioural outcome of purchase intention. To isolate the psychological effects of XR, we intentionally refrained from subdividing the CAB model beyond its established components and instead treated participants' reactions as composite internal responses. Although the CAB model formally distinguishes between cognitive and affective components, we did not impose a causal ordering between comprehension and empathy. Rather, we conceptualised both as higher-order internal reactions elicited by the respective advertising modalities and measured at a granularity appropriate for applied evaluation. Participants viewed identical advertising content across the two device conditions and rated the extent to which their comprehension of the product deepened, their empathy increased, and their inclination to purchase the product changed. This design enables us to clarify how immersive XR experiential advertising shapes internal responses, how these responses inform behavioural intentions, and how the CAB model may be applied to articulate the psychological impact of XR in advertising contexts.

## 2. Methods

### 2.1. Participants

Participants were recruited through a staffing agency and through announcements posted on the website of MESON Inc. All participants were required to have no financial relationship with the authors' organisation and not to be affiliated with any competing company. Screening criteria specified that participants must be healthy adults aged between 23 and 49 years, possess normal or corrected-to-normal vision of at least 1.0 using either unaided sight or contact lenses, report no colour-vision deficiencies or presbyopia, have no history of psychiatric disorders, and not be employed in the advertising or marketing industries.

Eighteen individuals met these criteria and took part in the study (6 men and 12 women; $M = 31.1 \pm 7.3$ years old). To ensure that the sample did not include participants with unusually high affinity for XR, we administered the Affinity for Technology Interaction (ATI) scale (Franke *et al.*, 2019) after the experiment. The mean ATI score was 4.26 (SD = 0.77). A Shapiro-Wilk test confirmed the absence of outliers ($w = 0.952$, $p = .460$).

Informed consent was obtained from all participants prior to participation. Compensation was provided in accordance with the organisation's internal hourly-rate guidelines.

### 2.2. Experimental Conditions

The experiment employed two within-participant conditions: a non-immersive 2D viewing advertising condition presented on a flat-panel display and an immersive XR experiential advertising condition delivered through a head-mounted display. The two setups are illustrated in Figure 1.

In the FPD condition, participants were seated approximately 2,000 mm from a 55-inch flat-panel display (KL-55X85J, Sony; width 1,119 mm, height 649 mm). The display subtended a horizontal visual angle of 31.5° and a vertical visual angle of 18.6°. Video stimuli were presented using QuickTime Player on a MacBook Air (Apple M2, 2022) connected via HDMI.

In the HMD condition, participants experienced the advertising content through an XR application built for the Apple Vision Pro (1st generation, 2024). Although the field of view is not publicly disclosed, estimates suggest approximately 100° horizontally and 80° vertically. Participants were seated facing the opposite direction from the FPD condition to avoid overlap between conditions, and they viewed the XR content against a white wall within a controlled space of approximately 4.0 m².

The FPD condition was designed to approximate a conventional television-viewing environment. In contrast, the HMD condition afforded interactive capabilities, enabling participants to lean in

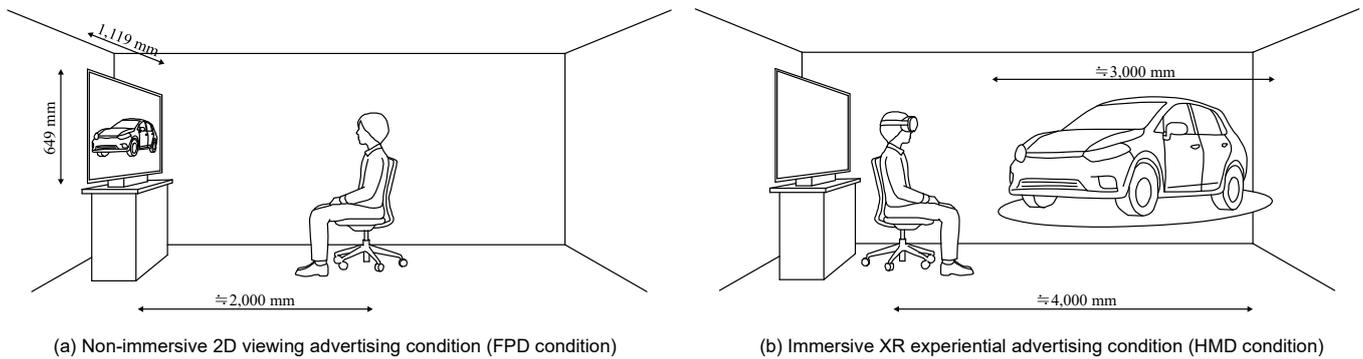

(a) Non-immersive 2D viewing advertising condition (FPD condition)   (b) Immersive XR experiential advertising condition (HMD condition)

**Figure 1. Viewing environments for the experimental stimuli.** (a) In the FPD condition, participants viewed the advertisements on a flat-panel display positioned approximately 2,000 mm in front of them. (b) In the HMD condition, participants faced the opposite direction, away from the flat-panel display, and viewed the XR advertisements superimposed within an approximately 4.0 m² area centred in front of them.

and inspect regions of the 3D model from different viewpoints, and further provided an augmented-reality overlay anchored to the physical testing space.

## 2.3. Stimuli

Twelve stimuli were prepared in total, comprising six for the FPD condition and six for the HMD condition (A-F in each; Table 1).

### 2.3.1 3D Models

The 3D models were selected from two industry domains in which XR advertising is considered highly applicable: automotive and housing/architecture. Three models were chosen from each domain. These industries typically incur high costs for maintaining physical showrooms or model houses and often face practical constraints in presenting multiple physical products to consumers. Consequently, they represent plausible use cases for XR-based advertising.

All models were pre-existing assets purchased from commercial 3D model marketplaces. To avoid brand-related bias, no models replicated identifiable products from specific manufacturers. Minor visual adjustments, such as rendering modifications, were applied where necessary to ensure stylistic consistency across stimuli.

### 2.3.2 XR Content and 2D Videos

The XR stimuli were developed using a research-purpose prototype XR application (*see* Footnote).

For the HMD condition, each stimulus consisted of a single sequence. The sequence began with a scaled-down exterior view of the model positioned within the participant's front-facing 4.0 m² physical space (AR display, approximately 10 seconds). Participants were then transitioned seamlessly into interior scenes (VR display, approximately 40 seconds) using skybox-based 360-degree environmental imagery. Finally, the sequence returned to an exterior view (AR display, approximately 10 seconds). All transitions were implemented as continuous walkthroughs to ensure a coherent experience. Automotive stimuli shared an identical camera perspective, while architectural stimuli were aligned as closely as possible in viewpoint. All XR content was created using the Unity PolySpatial SDK and built for the Apple Vision Pro at 90 fps.

**Table 1. Details of the 3D models used as experimental stimuli.** Models were selected from the automotive and housing/real-estate domains, which are considered promising domains for XR-based advertising, and representative content from each domain was included.

| Stimuli | A | B | C | D | E | F |
|---|---|---|---|---|---|---|
| Domain | | Automotive | | | Housing / Architecture | |
| Model | Sedan | Compact car | SUV | RC house | High-rise apartment | Timber house |

The 2D video stimuli for the FPD condition were generated from the corresponding XR sequences using Unity's virtual camera system and exported as MP4 files.

### 2.3.3 Audio Narration

Each stimulus was accompanied by an audio narration describing general product characteristics. Scripts were generated using a large language model (GPT-5) and converted into speech using text-to-speech software, then synchronised with the visual content. To avoid introducing brand-related biases, narrations refrained from highlighting specific product advantages and instead employed abstract promotional language.

## 2.4. Procedure

The experiment took place in a controlled laboratory environment within the researchers' organisation, with lighting and ambient noise kept constant.

Participants completed two blocks: one beginning with the FPD condition and the other with the HMD condition. Block order was counterbalanced across participants ($n = 9$). Each block comprised six trials. Each trial involved viewing one advertisement (approximately 60 seconds) followed by evaluative ratings. Stimulus order was randomised using a Latin square.

After each advertisement, participants rated three items on an 11-point scale from 0 ("not at all") to 10 ("very much"):

(1) the extent to which their comprehension of the product deepened (comprehension)

(2) the extent to which their empathy towards the product increased (empathy)

(3) the extent to which they felt inclined to purchase the product (purchase intention)

It is important to note that this measure of comprehension reflects the participants' subjective assessment of their understanding (i.e., perceived comprehension), rather than an objective test of knowledge retention. Participants were instructed to consider not whether they would actually purchase the product, but the extent to which the advertisement itself increased their purchase motivation.

Before the main experiment, participants completed a practice session in which they viewed a cruising-experience advertisement unrelated to the main stimuli in both conditions. This session allowed participants to familiarise themselves with the devices and the rating interface. After the first six-trial block, a short rest period (approximately five minutes) was provided before completing the second block. A brief post-experimental questionnaire followed.

All procedures complied with the Japanese Psychological Association's Ethical Principles (3rd edition, 2011) and were approved through the organisation's internal ethics process. Participants were informed that the experiment would be terminated immediately if symptoms of VR-induced discomfort occurred.

## 2.5. Analysis

*A priori* power analysis indicated that a minimum of 18 participants was required to detect a large effect size (Cohen's $d$ = 0.8) using G*Power 3.1 (Faul *et al*., 2007; Faul *et al*., 2009).

Analyses were conducted in Python using the statsmodels library. A repeated-measures two-way ANOVA was performed with advertising modality (FPD vs. HMD) and evaluation item (comprehension, empathy, purchase intention) as within-participant factors. Assumptions of normality and homogeneity of variance were confirmed prior to analysis. When significant main effects or interactions were observed, pairwise comparisons were conducted with Bonferroni correction.

Following Montoya and Hayes (2017), a within-participant mediation analysis was conducted. For each variable, a difference score was calculated by subtracting the FPD score from the HMD score ($\Delta$comprehension, $\Delta$empathy, and $\Delta$purchase intention). Comprehension and empathy served as mediators, and purchase intention served as the dependent variable in a parallel mediation model. The $\Delta$-values met normality assumptions (Shapiro-Wilk test) and showed no multicollinearity; therefore, mediation analysis used percentile bootstrapping (5,000 resamples) with $N$ = 18.

**Table 2. Results of the two-way repeated-measures ANOVA.** A significant main effect of advertising modality was observed, $F_{(1, 17)}$ = 177.42, $p < .001$. No significant main effect of evaluation item or interaction was detected, $F_{(2, 34)}$ = 0.97, $p = .384$; $F_{(2, 34)}$ = 0.35, $p = .708$.

| Factor | SS | F | P | Partial $\eta^2$ |
|---|---|---|---|---|
| Conditions | 132.963 | 177.42 | < 0.001 | .676 |
| Evaluation items | 1.449 | 0.967 | 0.384 | .022 |
| Interaction | 0.519 | 0.346 | 0.708 | .008 |
| Error | 63.703 | — | — | — |
| Total | 198.634 | — | — | — |

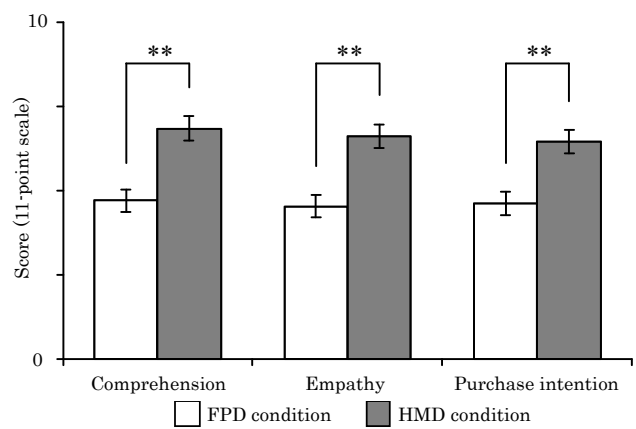

**Figure 2. Results for the three evaluation items across the two presentation conditions.** Paired-samples *t*-tests were conducted to follow up the significant main effect of condition: comprehension, $t_{(17)}$ = 8.92, $p < .001$, $dz$ = 2.10; empathy, $t_{(17)}$ = 7.06, $p < .001$, $dz$ = 1.66; purchase intention, $t_{(17)}$ = 5.81, $p < .001$, $dz$ = 1.37. All differences remained significant after Bonferroni correction (all corrected $p < .001$).

Effect sizes were reported as partial $\eta^2$ for the ANOVA, interpreted in line with Cohen (1988). In the mediation analysis, unstandardised coefficients ($B$) and their 95% confidence intervals were reported to evaluate effect magnitude and precision.

## 3. Results

### 3.1. Differences in Evaluation Scores Across Conditions

Table 2 presents the results of the two-way repeated-measures ANOVA examining the effects of advertising modality (FPD vs. HMD) and evaluation item (comprehension, empathy, purchase intention). Figure 2 displays the corresponding mean scores, with error bars representing standard errors.

There was a significant main effect of advertising modality, indicating that the immersive XR experiential advertising condition yielded consistently higher scores than the non-immersive 2D viewing advertising condition, $F_{(1, 17)}$ = 177.42, $p < .001$. Inspection of the magnitude of score increases revealed that comprehension rose by 45.4 %, empathy by 45.6 %, and purchase intention by 39.9 % in the HMD condition compared with the FPD condition. The increases in comprehension and empathy, both internal responses, were therefore relatively larger than the increase in the behavioural outcome of purchase intention.

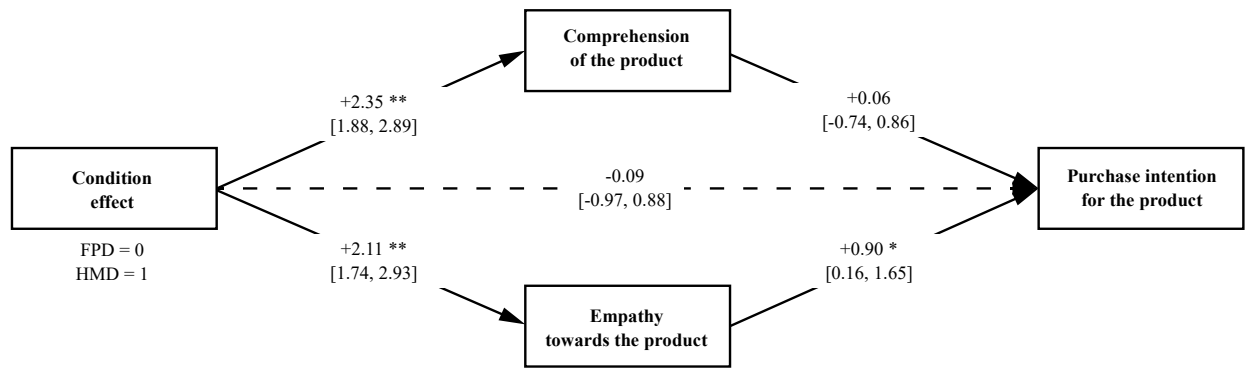

**Figure 3. Parallel mediation analysis for comprehension and empathy.** The indirect path through comprehension was not significant ($B = +0.13$, 95% CI [−1.68, 2.10], *n.s.*), whereas the indirect path through empathy was significant ($B = +2.09$, 95% CI [0.33, 3.93], $p < .05$). The combined indirect effect across both mediators was significant ($B = +2.22$, 95% CI [1.31, 3.14], $p < .01$).

In contrast, there was no significant main effect of evaluation item, $F_{(2, 34)} = 0.97$, $p = .384$, and no significant interaction between advertising modality and evaluation item, $F_{(2, 34)} = 0.35$, $p = .708$. Mauchly's test was used to evaluate the assumption of sphericity, and since neither the main effect nor the interaction violated this assumption, no corrections were applied.

To further examine differences between conditions, paired-samples *t*-tests were performed for each evaluation item. Significant differences were observed for all three items: comprehension, $t_{(17)} = 8.92$, $p < .001$; empathy, $t_{(17)} = 7.06$, $p < .001$; and purchase intention, $t_{(17)} = 5.81$, $p < .001$. All differences remained significant following Bonferroni correction (all corrected $p < .001$).

### 3.2. Contribution of Internal Responses to Behavioural Intention

Figure 3 and Table 3 present the results of the parallel mediation analysis assessing how differences in internal responses, specifically comprehension and empathy, mediated the effect of advertising modality on purchase intention.

The total effect (*c*) of advertising modality was significant, indicating that immersive XR experiential advertising increased purchase intention relative to non-immersive 2D viewing advertising, $B = +2.13$, 95% CI [1.48, 2.76], $p < .001$. However, the direct effect (*c'*), controlling for the mediators, was not significant, $B = −0.09$, 95% CI [−0.97, 0.88].

Regarding indirect effects, the mediation pathway through comprehension was not significant, $B = +0.13$, 95% CI [−1.68, 2.10]. In contrast, the pathway through empathy was significant, $B = +2.09$, 95% CI [0.33, 3.93], $p < .05$. The combined indirect effect across both mediators was also significant, $B = +2.22$, 95% CI [1.31, 3.14], $p < .01$.

A decomposition of the model further showed that the HMD condition, compared with the FPD condition, significantly increased scores for comprehension, $B = +2.35$, 95% CI [1.88, 2.89], $p < .001$, and for empathy, $B = +2.31$, 95% CI [1.74, 2.93], $p < .001$. The pathway from empathy to purchase intention was significant, B = +0.90, 95% CI [0.16, 1.65], $p < .05$, whereas the pathway from comprehension to purchase intention was not, $B = +0.06$, 95% CI [−0.74, 0.86].

**Table 3. Mediation analysis results.** The total effect (*c*) was significant ($B = +2.13$, 95% CI [1.48, 2.76], $p < .001$). The direct effect (*c'*), controlling for the mediators, was not significant ($B = −0.09$, 95% CI [−0.97, 0.88], *n.s.*).

| Effect | | Unstandardised coefficient *B* | 95% confidence interval | *P* |
|---|---|---|---|---|
| Total effect (*c*) | | +2.13 | [1.48, 2.76] | < 0.001 |
| Direct effect (*c'*) | | -0.09 | [-0.97, 0.88] | *n.s.* |
| Indirect effects | Comprehension | +0.13 | [-1.68, 2.10] | *n.s.* |
| | Empathy | +2.09 | [0.33, 3.93] | <0.05 |
| Total indirect effect | | +2.22 | [1.31, 3.14] | <0.01 |

Taken together, these findings indicate that immersive XR experiential advertising enhanced both comprehension and empathy relative to non-immersive 2D viewing advertising. However, only empathy exerted a significant mediating influence on purchase intention, whereas comprehension did not. Thus, of the two hypothesised pathways whereby internal responses might shape behavioural intention, the empathy pathway was supported and the comprehension pathway was not.

## 4. Discussion

### 4.1 The Enhancement of Empathy and the Limited Role of Comprehension

This study examined how immersive XR experiential advertising, compared with non-immersive 2D viewing advertising on a flat-panel display, influences internal responses and behavioural intention. The results showed that the head-mounted display condition produced higher ratings across all three measures, namely comprehension, empathy, and purchase intention. Mediation analysis, however, demonstrated that only the pathway through empathy significantly predicted purchase intention, whereas the pathway through comprehension was not significant. These findings suggest that although XR advertising enhances internal responses as a whole, corresponding to the cognitive and affective components of the CAB model, the principal psychological driver of the behavioural component, purchase intention, is affective rather than cognitive. Moreover, the influence of XR on users' mental states appears to arise not from mere cognitive facilitation, but from a strengthening of affective engagement grounded in the experiential quality of the medium.

These results align with a substantial body of work indicating that immersive VR elicits heightened presence and enhances empathy through perspective-taking (Han *et al.*, 2022; Herrera *et al.*, 2018; Schutte and Stilinović, 2017). Meta-analytic evidence further indicates that VR-induced empathy varies depending on narrative structure and experiential design, thereby supporting the proposition that the mediational pathway identified in the present study, from XR to empathy to purchase intention, reflects a robust influence of XR's affective qualities on behavioural expression (Ventura *et al.*, 2020; Martingano *et al.*, 2021).

The observed enhancement of empathy can be interpreted in relation to the bodily and spatial affordances of XR. Affordances refer to the action possibilities offered by an environment (Gibson, 1979). Immersive VR affords users a sense of being located within the virtual environment and of being able to act within it as if physically present. From this perspective, the well-established capacity of VR to elicit place illusion and plausibility illusion can be viewed as providing the perceptual basis for actions such as approaching, leaning in, or visually exploring a scene (Slater and Sanchez-Vives, 2016). Embodiment research similarly suggests that active viewpoint shifts and physical proximity within virtual environments reduce psychological distance and strengthen emotional attachment. In the present study, the HMD condition enabled participants to inspect products at close range and freely direct their gaze, fostering an experience of active perception. This likely accentuated subjective engagement relative to 2D viewing and strengthened affective connection. These considerations imply that XR, by virtue of its inherent affordances, is predisposed to amplify the affective component of the CAB model.

By contrast, the cognitive pathway of the CAB model, namely comprehension, did not significantly mediate purchase intention. This is likely attributable to the design choices of the present stimuli, which intentionally excluded specific cognitive-support features in order to avoid brand bias. Although immersive VR can support learning and knowledge acquisition (Merchant *et al.*, 2014), such benefits depend heavily on the appropriateness of task design and attentional guidance (Mikropoulos and Natsis, 2011). In the present study, narration was deliberately abstract and did not highlight specific product attributes. In an XR environment that affords free exploration, the absence of explicit cues may have dispersed attention, thereby attenuating the operation of the comprehension pathway.

This interpretation is consistent with established findings in multimedia learning, including evidence that visual and auditory information jointly enhance comprehension (Mayer, 2001), that verbal and imagery codes jointly improve retention and comprehension according to dual-coding theory (Paivio, 1986), and that auditory cues can direct attention through orienting responses (Lang, 2000). Research further shows that visual and verbal guidance facilitates comprehension (Jamet *et al.*, 2008; de Koning *et al.*, 2010) and that congruence between auditory and visual information enhances memory and semantic interpretation (Heckler and Childers, 1992; Kong *et al.*, 2019). The XR stimuli in the present study had strong visual immersion but relatively weak cognitive support, which likely limited the contribution of the cognitive component of the CAB model.

Taken together, the results indicate that among the internal responses elicited by XR, the affective component plays a particularly prominent role in shaping behavioural intention. Nevertheless, to situate these findings within the broader context of advertising, it is necessary to consider their implications from a more applied perspective.

### 4.2. XR in Advertising Contexts

To contextualise the present findings within advertising practice, it is essential to reconsider the dual role of XR advertising as both a medium capable of eliciting affective immersion and a medium that can, under appropriate conditions, facilitate information-based comprehension. The non-significant comprehension pathway observed in this study is not indicative of a structural limitation of XR. Rather, it reflects the absence of multimodal guidance, which was intentionally minimised to avoid brand-related bias. XR experiences readily capture attention through visual immersion, yet effective comprehension requires cognitive scaffolding, such as auditory guidance, visual highlighting, or structured sequencing that links perceptual exploration to meaning-making. Given that these elements were deliberately attenuated, participants may have prioritised free exploration, resulting in insufficient gains in comprehension to influence purchase intention.

Accordingly, the non-significant comprehension pathway does not undermine XR's potential as a medium for facilitating comprehension. Instead, it suggests that such potential is highly malleable and dependent on the design of appropriate guidance. By incorporating staged narration, interactive manipulation, or emphasising salient internal features of a product, XR advertising can be configured to evoke substantial cognitive as well as affective responses.

When considering the psychological effects of XR advertising, it is also important to acknowledge its inherently embodied nature. XR enables users to engage with content through active visual and bodily movements, fostering a sense of "being there" that exceeds what can be achieved through conventional visual media. Phenomenological accounts, particularly those of Merleau-Ponty, highlight that perception arises through embodied interaction with the world; this embodied embeddedness influences attitude formation. The present findings likewise suggest that embodied XR experiences enhance affective engagement, thereby shaping behavioural intention. This supports the view that XR holds distinct advertising value grounded in its capacity to elicit embodied presence.

From the standpoint of advertising practice, the strong affective engagement characteristic of XR represents a significant advantage for brand storytelling and experiential marketing. However, for advertisements that prioritise functional information or technical specifications, excessive reliance on free exploration may diffuse the focus of information delivery. In such cases, designs that combine visual and auditory cues to guide attention become essential. Moreover, to gain a more comprehensive understanding of the psychological effects of XR advertising, future research should integrate a wider set of psychological constructs, including presence, self-involvement, enjoyment, trust, and curiosity, alongside comprehension and empathy. Building such an integrative model would allow a more precise explanation of the mechanisms through which XR advertising exerts its influence.

This study has several limitations. Given the relatively small sample size ($N = 18$), the present study should be interpreted as a pilot investigation. Specifically regarding the mediation analysis, further research with a larger sample size is required to achieve robust statistical power. Additionally, the experimental stimulus design warrants further refinement to rigorously disentangle the effects of cognitive and affective guidance. Other limitations include the abstract nature of the stimuli, the limited range of product categories, the single-exposure design, and reliance on self-report measures. Additionally, the study assessed short-term evaluation, leaving the long-term effects of XR experiences on attitudes and actual purchasing behaviour unresolved. Future work should incorporate systematic manipulations of guidance features, eye-tracking-based analyses of attention, logging of in-experience behaviour, longitudinal follow-up, and comparisons across product types. By appropriately designing and leveraging both the experiential and informational affordances of XR, advertisers may create novel forms of advertising that surpass those afforded by conventional media.

## 5. Conclusion

The present study demonstrated that immersive XR experiential advertising delivered through a head-mounted display elicits higher levels of purchase intention than conventional non-immersive 2D viewing advertising presented on a flat-panel display. Furthermore, examination of the internal contributors to behavioural intention showed that differences in the viewing experience influenced purchase intention primarily through the mediating role of empathy. In contrast, comprehension did not show a statistically significant mediating effect in this experimental context. This pattern is consistent with prior research reporting that XR experiences enhance empathy. By advancing our understanding of both the experiential and informational affordances of XR, future work may help articulate design principles for more effective XR-based advertising experiences.

### Footnote

The XR application used in this study was an internal prototype developed for research purposes and is not a consumer-facing product (informally referred to as the "*Immersive Showroom*").

### Acknowledgements

The authors thank ANDO Masahito for his support in producing the experimental stimuli and implementing the application on the Apple Vision Pro, and SAKAMOTO Hiroyuki for his assistance in creating several of the stimuli. The direction of the experimental advertising stimuli was carried out with the support of internal colleagues.

### Author Contributions

YK developed the research hypotheses. KT designed and conducted the experiment, performed the data analysis, contributed to the interpretation, and wrote the manuscript. Both authors contributed to the final manuscript and approved the submitted version.

### Conflict of Interest

This research was conducted as part of the authors' duties at MESON Inc. The first author is the company's Chief Executive Officer, and the second author is an employee. No additional conflicts of interest related to funding or study execution are declared.